\documentclass[12pt,a4paper]{article}
\usepackage{amsfonts}
\usepackage{amsfonts}
\usepackage{amsfonts}
\usepackage{amsfonts}
\usepackage{amsfonts}
\usepackage{amsfonts}
\usepackage{amsfonts}
\usepackage{amsfonts}
\usepackage{amsfonts}
\usepackage{amsfonts}
\usepackage{amsfonts}
\usepackage{amsfonts}
\usepackage{mathrsfs}
\usepackage{mathrsfs}
\usepackage{mathrsfs}
\usepackage{mathrsfs}
\usepackage{mathrsfs}
\usepackage{mathrsfs}
\usepackage{mathrsfs}
\usepackage{mathrsfs}
\usepackage{mathrsfs}
\usepackage{amsfonts}
\usepackage{amsfonts}
\usepackage{amsfonts}
\usepackage{amsfonts}
\usepackage{amsfonts}
\usepackage{amsfonts}
\usepackage{mathrsfs}
\usepackage{amsfonts}
\usepackage{amsfonts}
\usepackage{amsfonts}
\usepackage{amsfonts}
\usepackage{amsfonts}
\usepackage{amsfonts}
\usepackage{amsfonts}
\usepackage{amsfonts}
\usepackage{amsfonts}
\usepackage{amsfonts}
\usepackage{amsfonts}
\usepackage{amsfonts}
\usepackage{mathrsfs}
\usepackage{amsfonts}
\usepackage{mathrsfs}
\usepackage{amsfonts}
\usepackage{amsfonts}
\usepackage{amsfonts}
\usepackage{mathrsfs}
\usepackage{amsfonts}
\usepackage{amsfonts}
\usepackage{amsfonts}
\usepackage{amsfonts}
\usepackage{amsmath}
\usepackage{mathrsfs}
\usepackage{amsfonts}
\usepackage{amsfonts}
\usepackage{amsfonts}
\usepackage{amsfonts}
\usepackage{amsfonts}
\usepackage{amsfonts}
\usepackage{amsfonts}
\usepackage{amsfonts}
\usepackage{amsfonts}
\usepackage{amsfonts}
\usepackage{amsfonts}
\usepackage{mathrsfs}
\usepackage{amsfonts}
\usepackage{amsfonts}
\usepackage{amsfonts}
\usepackage{amsfonts}
\usepackage{amsmath}
\usepackage{mathrsfs}
\usepackage{mathrsfs}
\usepackage{mathrsfs}
\usepackage{mathrsfs}
\usepackage{mathrsfs}
\usepackage{mathrsfs}
\usepackage{mathrsfs}
\usepackage{mathrsfs}
\usepackage{amsfonts}
\usepackage{amsfonts}
\usepackage{amsfonts}
\usepackage{amsfonts}
\usepackage{amsfonts}
\usepackage{amsfonts}
\usepackage{amsfonts}
\usepackage{amsfonts}
\usepackage{amsfonts}
\usepackage{amsfonts}
\usepackage{amsfonts}
\usepackage{amsfonts}
\usepackage{mathrsfs}
\usepackage{amsfonts}
\usepackage{mathrsfs}
\usepackage{amsfonts}
\usepackage{amsfonts}
\usepackage{amsfonts}
\usepackage{amsfonts}
\usepackage{amsfonts}
\usepackage{amsfonts}
\usepackage{amsfonts}
\usepackage{amsfonts}
\usepackage{amsfonts}
\usepackage{amsfonts}
\usepackage{amsfonts}
\usepackage{amsfonts}
\usepackage{amsfonts}
\usepackage{amsfonts}
\usepackage{amsfonts}
\usepackage{amsfonts}
\usepackage{amsfonts}
\usepackage{amsfonts}
\usepackage{mathrsfs}
\usepackage{mathrsfs}
\usepackage{mathrsfs}
\usepackage{mathrsfs}
\usepackage{amsfonts}
\usepackage{amsfonts}
\usepackage{amsfonts}
\usepackage{amsfonts}
\usepackage{amsfonts}
\usepackage{amsfonts}
\usepackage{amsfonts}
\usepackage{amsfonts}
\usepackage{amsfonts}
\usepackage{amsfonts}
\usepackage{amsfonts}
\usepackage{amsfonts}
\usepackage{amsfonts}
\usepackage{amsfonts}
\usepackage{amsfonts}
\usepackage{amsfonts}
\usepackage{amsfonts}
\usepackage{mathrsfs}
\usepackage{mathrsfs}
\usepackage{amsfonts}
\usepackage{amsfonts}
\usepackage{amsfonts}
\usepackage{amsfonts}
\usepackage{amsfonts}
\usepackage{amsfonts}
\usepackage{amsfonts}
\usepackage{amsfonts}
\usepackage{amsfonts}
\usepackage{amsfonts}
\usepackage{amsfonts}
\usepackage{amsfonts}
\usepackage{amsfonts}
\usepackage{mathrsfs}
\usepackage{mathrsfs}
\usepackage{amsfonts}
\usepackage{amsfonts}
\usepackage{amsfonts}
\usepackage{amsfonts}
\usepackage{amsfonts}
\usepackage{amsfonts}
\usepackage{mathrsfs}
\usepackage{mathrsfs}
\usepackage{amsfonts}
\usepackage{amsfonts}
\usepackage{amsfonts}
\usepackage{amsfonts}
\usepackage{amsfonts}
\usepackage{amsfonts}
\usepackage{amsfonts}
\usepackage{amsfonts}
\usepackage{amsfonts}
\usepackage{amsfonts}
\usepackage{amsfonts}
\usepackage{amsfonts}
\usepackage{amsfonts}
\usepackage{amsfonts}
\usepackage{mathrsfs}
\usepackage{amsfonts}
\usepackage{amsfonts}
\usepackage{amsfonts}
\usepackage{amsfonts}
\usepackage{amsfonts}
\usepackage{amsfonts}
\usepackage{amsfonts}
\usepackage{amsfonts}
\usepackage{amsfonts}
\usepackage{amsfonts}
\usepackage{amsfonts}
\usepackage{amsfonts}
\usepackage{amsfonts}
\usepackage{amsfonts}
\usepackage{amsfonts}
\usepackage{amsfonts}
\usepackage{amsmath}
\usepackage{mathrsfs}
\usepackage{amsfonts}
\usepackage{amsfonts}
\usepackage{amsfonts}
\usepackage{amsfonts}
\usepackage{amsmath}
\usepackage{mathrsfs}
\usepackage{amsfonts}
\usepackage{mathrsfs}
\usepackage{mathrsfs}
\usepackage{mathrsfs}
\usepackage{mathrsfs}
\usepackage{amsmath}
\usepackage{mathrsfs}
\usepackage{amsfonts}
\usepackage{color}
\usepackage{mathrsfs}
\usepackage{setspace}

\usepackage{stmaryrd}
\usepackage{amsmath}
\usepackage{amssymb}

\usepackage{bbm}
\usepackage{mathrsfs}
\usepackage{graphicx}
\usepackage{cite}
\usepackage[colorlinks, linkcolor=blue, anchorcolor=blue, citecolor=blue]{hyperref}
\usepackage{amssymb}
\usepackage{enumerate}
\usepackage{theorem}
\makeatletter
\def\tank#1{\protected@xdef\@thanks{\@thanks
        \protect\footnotetext[0]{#1}}}
\def\bigfoot{

    \@footnotetext}
\makeatother

\newtheorem{thm}{Theorem}[section]
\newtheorem{lem}[thm]{Lemma}

\newtheorem{rem}[thm]{Remark}

\newtheorem{defn}[thm]{Definition}

\newenvironment{proof}{Proof.}

\renewcommand{\d}{d}

\def\ba{\begin{array}}
\def\ea{\end{array}}
\def\beq{\begin{equation}}
\def\bes{\begin{equation*}}
\def\ees{\end{equation*}}
\def\bea{\begin{eqnarray}}
\def\eea{\end{eqnarray}}
\def\beaa{\begin{eqnarray*}}
\def\eeaa{\end{eqnarray*}}

\def\nts{\negthinspace}

\def\={=\nts \nts=\nts \nts=\nts \nts=}

\def\bh{\mbox}
\def\({\textnormal{(}}
\def\){\textnormal{)}}

\def\qx{\wedge}

\def\qx{\wedge}

\def\g{\gamma}
\def\d{\delta}
\def\e{\varepsilon}

\def\l{\lambda}

\def\t{\tau}
\def\yj{\hat}
\def\yw{\tilde}

\def\f{\phi}
\def\vf{\varphi}
\def\p{\psi}

\def\kspl{\begin{itemize}}
\def\jspl{\end{itemize}}

\def\wq{\infty}

\def\bsy{\notin}

\def\hx{\mathcal}
\def\yjt{\rightarrow}

\def\jf{\int}
\def\fs{\frac}
\def\bsy{\notin}
\def\ksfl{\begin{cases}}
\def\jsfl{\end{cases}}

\def\b1{{\bf 1}}

\begin{document}
\title{\Large \bf Optimal Dividend  of  Compound Poisson Process under a Stochastic Interest Rate
\thanks{Research is supported by Chinese NSF Grants No.11471171 and No.11571189.} }
\author{{Linlin Tian}$^a$\footnote{E-mail:tianlinlin1992@163.com}~~~{Xiaoyi Zhang}$^b$\footnote{Corresponding author, E-mail: zhangxiaoyi19902@163.com}~~~
\\
\\
 \small  a. School of Mathematical Sciences, Nankai University, Tianjin 300071, China.\\
\small  b. School of Mathematical Sciences, Nankai University, Tianjin 300071, China.}\,
\date{}
\maketitle
\begin{center}
\begin{minipage}{130mm}
{\bf Abstract} \;\;In this paper we assume the insurance wealth process is driven by the compound Poisson
process. The discounting factor is modelled as a  geometric Brownian motion at first and then as an exponential function of
an integrated Ornstein-Uhlenbeck process. The objective is to maximize the cumulated value
of expected discounted dividends up to the time   of ruin. We give an explicit expression of the
value function and the optimal strategy in the case of  interest rate following a geometric Brownian motion. For the  case of  the Vasicek model, we explore some
properties of the value function. Since we can not find an explicit expression for the value function in
the second case, we prove that the value function is
the viscosity solution of the corresponding HJB equation.

\vspace{3mm} {\bf Keywords:} Hamilton-Jacobi-Bellman equation, Vasicek model, Geometric Brownian motion,
Interest rate, Viscosity solution, Optimal dividends

\end{minipage}
\end{center}

\section{Introduction}
The optimal dividend problem has been discussed for a long time in the literature. In 1957 De
Finetti  \cite{de1957impostazione} proposed that an insurance company should allow cash leakages and measure their performance during its life time instead of only focussing on  ruin probability. These cash  leakages can be interpreted as dividends.
In the setting of  constant interest rate, Asmussen and Taksar\cite{Asmussen1997Controlled} solved the optimal dividend problem for the special case of Brownian motion. They found out that the optimal strategy is a constant barrier strategy in the case of unbounded dividend and a so-called threshold strategy in the case of restricted dividend rates.  In the case of a  surplus process following a compound Poisson process, Gerber and Shiu \cite{gerber2006optimal} showed that  the optimal  strategy is a  threshold strategy when claim size are exponentially distributed for restricted dividend rates.
For the more general  case of claim size distribution, Azcue and Muler  \cite{azcue2005optimal} studied the optimal reinsurance and dividend  policy in the framework of Cram$\acute{\mbox{e}}$r-Lundberg model using viscosity solution.
Later, Azcue and Muler, see \cite{azcue2012optimal}, found the optimal dividend payment policy in the case of bounded dividend rates.
In the setting of constant interest rate the optimal dividend problem has been studied quite well under various general reserve models, see e.g.,\cite{albrecher2008optimal,loeffen2008optimality,schmidli2007stochastic}. We  omit listing the existing literature and  refer to a survey on the dividend problems by
Albrecher and Thonhauser \cite{albrecher2009optimality} and references therein.

The interest rate forms a key component of the financial market,  influencing the firm's cost and profit.   There are a lot of factors influencing interest rate, such as inflation rate, monetary policy, exchange rate policy, international agreement, and international privity.
The interest rate is also an important tool reflecting policy makers' intentions and achieving economic objectives.
As it changes over time, it  is more reasonable to assume that the interest rate is a function of time instead of a  deterministic constant. The changes of interest rate  reflect the fluctuations of the monetary market.
Eisenberg \cite{Eisenberg2015Optimal} solved  optimal dividends problem in the setting of surplus following a drifted
Brownian motion. The discounting factor is modelled as a stochastic process: at first as a geometric
Brownian motion, then as an exponential function of an integrated Ornstein-Uhlenbeck process. They found an explicit expression for the value function of the optimal strategy for both restricted  and unrestricted dividends in the case of geometric Brownian motion.

In our paper, we model the surplus process as a compound Poisson process.  In section 3,  we explore the dividend maximization problem under the Dothan model and find, similar to the case of  deterministic interest rate, that the optimal strategy does not change (compared to the Gerber-Shiu case) in its form, but  the parameters do.
In Section 4,  we consider  the Vasicek model, for which the short rate is defined as an Ornstein-Uhlenbeck process. Here, the situation  changed completely. It is not that easy  to calculate the return function of the corresponding strategy.
We explore the continuity of the value function but unfortunately we can not prove more regularity properties about the value function. It is natural to consider the problem in the framework of viscosity solutions.
 \section{Problem Formulation}
 In our paper, the reserve $X_t$ of an insurance company can be described by
\begin{align}\label{1}
X_t=x+ct-\sum_{k=1}^{N(t)}Y_k,
\end{align}
where $x\ge 0$ is the initial surplus, the constant $c>0$ is the premium rate,
 $N(t)$ is the Poisson process representing the frequency of the incoming claims,  $\{Y_i\}_{i=1}^\wq$
representing a sequence of independent, identically distributed (i.i.d.) random variables  with  distribution $G:\mathbb{R}^+\yjt \mathbb{R}$.
Assume that the insurance company is allowed to pay out dividends, where the accumulated dividends until time $t$ are given by $L_t$. The surplus at time $t$ is described as:
\begin{align}\nonumber
X_t^L=x+ct-\sum_{k=1}^{N(t)}Y_k-L_t.
\end{align}
Denote $B_t$ a standard Brownian motion.  All of the above defined quantities are defined on the same filtered probability space $(\Omega, \mathcal{F},(\mathcal{F}_t)_{t\ge 0},\mathbb{P})$, with $(\mathcal{F}_t)_{t\ge 0}$ the filtration generated by  $\{B_t,X_t\}$.
Here we only allow the restricted dividend, which means, the cumulative dividend up to time $t$ is given by $L_t=\jf_0^t l_sds$, with  $l_s\in [0,M]$ for some constant $M>0$. We say that a strategy $L$ is admissible if it is predictable, nondecreasing, cadlag and it verifies $X_t^L\ge 0$ up to the ruin time.   Denote $\mathcal{U}_{ad}$ the set of all admissible strategies.
Our target is to find the optimal  strategy maximizing the expectation of the cumulative discounted dividend under two different kinds of stochastic  interest rates. First, we  consider  a geometric Brownian motion model  and then we consider the Vasicek model.
\section{Geometric Brownian motion as a discounting factor}
In this section, we make the assumption that $M<c$ for mathematical convenience, which means the dividend rate  can not exceeds the premium rate.
We also specify $G(x)=1-e^{-\beta x}$, which means  claims follow an exponential distribution with rate $\beta>0$.
As a risk measure, we consider that dividends are discounted by the geometric Brownian motion
\begin{align}\nonumber
\exp\{-r-mt-\d B_t\}.
\end{align}
Here we denote $r_t=r+mt+B_t$ with initial value $r$.
Denoting by $\t^L$ the ruin time of the surplus process under some admissible strategy $L = \{l_s\}$, we define the return function corresponding to $L$ to be
\begin{align}
J^L(r,x)=E\left[\jf_0^{\t^L}e^{-r-ms-\d B_s}l_sds\right].
\end{align}
The objective is to  find an optimal  dividend policy $L_t$ to maximize the expectation of  cumulative discounted dividends. We denote  $V(r,x)$ the
optimal value function
\begin{align}
V(r,x)=\sup_{L\in \hx{U}_{ad}}J^L(r,x),
\end{align}
here $\hx{U}_{ad}$ denotes the set of all admissible strategies.
We note that for any strategy $L$,
\begin{align}\nonumber
J^L(r,x)=E\left[\jf_0^{\t^L}e^{-r-ms-\d B_s}l_sds\right]\le E\left[\jf_0^{\t^L}e^{-r-ms-\d B_s}Mds\right]
=\fs{Me^{-r}}{m-\fs{\d^2}{2}}.
\end{align}
This means $V(r,x)$ is bounded.
The HJB equation corresponding to  the problem is
\begin{align}
\label{HJB}
\left[mV_r+\fs{\d^2}{2}V_{rr}+cV_x-\l V \right](r,x)+\l \jf_0^xV(r,x-y)\beta e^{-\beta y}dy +\max_{l\in[0,M]}(e^{-r}-V_x(r,x))l=0.
\end{align}
\subsection{Solving HJB Equation}
Now we focus on solving the HJB equation. Denote $\mathbb{C}^1(\mathbb{R}^+)$ the set of all continuously differentiable function on $\mathbb{R}^+$.
We conjecture that $V(r,x)=e^{-r}F(x)$. We only need to find a function  $F(x)\in \mathbb{C}^1(\mathbb{R}^+)$ such that $F(x)$ satisfies
\begin{align}
\label{Fx}
\left(-m+\fs{\d^2}{2}-\l \right)F(x)+cF'(x)+\l\jf_0^xF(x-y)\beta e^{-\beta y}dy+\max_{l\in[0,M]}(1-F'(x))l=0.
\end{align}
We suppose that there exists a  concave function $F(x)$  satisfying equation (\ref{Fx}).  Because of the   linearity in the control $l$, we get a critical point $b^*$ with $F'(x)>1$ for $x<b^*$,  $F'(b^*)=1$ and $F'(x)<1$ for $x>b^*$. It is possible that $b^*=0$. Under these assumptions the HJB equation (\ref{Fx}) becomes
\begin{align}\label{Fx1}
&\left(-m+\fs{\d^2}{2}-\l\right )F(x)+cF'(x)+\l\jf_0^xF(x-y)\beta e^{-\beta y}dy=0, \quad 0<x<b^*;\\\label{Fx2}
&\left(-m+\fs{\d^2}{2}-\l\right )F(x)+cF'(x)+\l\jf_0^xF(x-y)\beta e^{-\beta y}dy+ M\left(1-F'(x)\right)=0,x\ge b^*.
\end{align}
Equation (\ref{Fx1}) can be written as
\begin{align}
cF''(x)+\left(\beta c-\l-\left(m-\fs{\d^2}{2}\right)\right)F'(x)-\beta\left(m-\fs{\d^2}{2}\right)F(x)=0,\; 0<x<b^*
\end{align}
with a general solution of the form
\begin{align}\label{gb1}
F(x)\triangleq F_1(x)=C_1 e^{R_1x}+C_2 e^{R_2x},
\end{align}
where $R_1>0$ and $R_2<0$ are the roots of the characteristic equation
\begin{align}\nonumber
c\xi^2+\left[\beta c-\l-\left(m-\fs{\d^2}{2}\right)\right]\xi-\beta\left(m-\fs{\d^2}{2}\right)=0.
\end{align}
Similarly, for all $x>b^*$, equation (\ref{Fx2}) can be written as
\begin{align}\label{cM}
(c-M)F''(x)+\left[\beta (c-M)-\l-\left(m-\fs{\d^2}{2}\right)\right]F'(x)-\beta\left(m-\fs{\d^2}{2}\right)F(x)+\beta M=0.
\end{align}
Combining (\ref{cM}) with the fact that $F(x)\le \fs{M}{m-\fs{\d^2}{2}}$, we know that (\ref{cM}) has a solution of the form
\begin{align}\label{fxx}
\forall \;x\ge b^*,\;\; F(x)\triangleq F_2(x)=\fs{M}{m-\fs{\d^2}{2}}+De^{S_2x},
\end{align}
where $D\le 0$ is a constant, and $S_2$ denotes the negative root of the following equation
\begin{align}\nonumber
(c-M)\xi^2+\left[\beta(c-M)-\l-(m-\fs{\d^2}{2})\right]\xi-\beta\left(m-\fs{\d^2}{2}\right)=0.
\end{align}
It is possible that $b^*=0$. If $b^*=0$, then (\ref{fxx}) satisfies (\ref{Fx2}) for all initial capital $x\ge 0$. Putting (\ref{fxx}) into (\ref{Fx2}) implies that (\ref{Fx2}) has a solution  of the form
\begin{align}\nonumber
F_2(x)=\fs{M}{m-\fs{\d^2}{2}}\left[1-e^{S_2x}\left(1+\fs{S_2}{\beta}\right)\right].
\end{align}
This function is increasing and concave because $\beta+S_2>0$. If $(-S_2)\fs{M}{m-\fs{\d^2}{2}}(1+\fs{S_2}{\beta})\le 1$, then $F_2'(0)\le 1$, in this case $F_2'(x)\le 1$ for all $x\ge 0$ and $F_2(x)$ is the solution of (\ref{Fx}). We can see that  $e^{-r}F_2(x)$ is the solution of HJB equation (\ref{HJB}).

From now on we consider the opposite case $(-S_2)\fs{M}{m-\fs{\d^2}{2}}(1+\fs{S_2}{\beta})>1$. We need to find a differentiable solution of (\ref{Fx1}) and (\ref{Fx2}).
Substituting (\ref{gb1}) into (\ref{Fx1}) and setting the coefficient of $e^{-\beta x}$ with $0$, we obtain that there exists a constant $\g>0$ ($\g$ is independent of $x$) such that
\begin{align}\nonumber
F_1(x)=\gamma[(R_1+\beta)e^{R_1x}-(R_2+\beta)e^{R_2x}],\;\;0\le x\le b^*.
\end{align}
From the continuity of $F(x)$ at $b^*$, which means $F_1(b^*)=F_2(b^*)$, we obtain
\begin{align}\label{condition1}
\g [(R_1+\beta)e^{R_1b^*}-(R_2+\beta)e^{R_2b^*}]=\fs{M}{m-\fs{\d^2}{2}}+De^{S_2b^*}.
\end{align}
Substituting (\ref{condition1}) and (\ref{fxx}) into (\ref{Fx2}), setting the coefficient of $e^{-\beta x}$ to $0$, and cancelling the factor $\beta e^{-\beta b^*}$, we obtain
\begin{align}\label{condition2}
\g(e^{R_1b^*}-e^{R_2b^*})-\fs{M}{\beta(m-\fs{\d^2}{2})}-\fs{De^{S_2b^*}}{\beta+S_2}=0.
\end{align}
To determine $\g, D, b^*$, we can use the condition
\begin{align}\label{condition3}
F_1'(b^*-)=F_2'(b^*+)=1.
\end{align}
Combining (\ref{condition1}), (\ref{condition2}) and (\ref{condition3}), we can obtain  closed-form expressions  for $\g, D, \bh{and} \;b^*$. It is not hard to see that $F_1'(x)>1$ on $(0,b^*)$ and  $F_2'(x)\le1$ on $[b^*,+\wq)$, we  omit the details here. As a summary, we  give out the following theorem.
\begin{thm} The solution of HJB equation (\ref{HJB}) is organised as follows.
%
%

\noindent
If $(-S_2)\fs{M}{m-\fs{\d^2}{2}}(1+\fs{S_2}{\beta})\le 1$,
\begin{align}\nonumber
V(r,x)=e^{-r}\fs{M}{m-\fs{\d^2}{2}}\left[1-e^{S_2x}\left(1+\fs{S_2}{\beta}\right)\right],
\end{align}
and the optimal dividend strategy $L^*=\{l_s^*\}$ is
\begin{align}\nonumber
l_s^*=M\mathbf{1}_{\{X_s^{L^*}\ge 0\}}.
\end{align}
 \noindent
If  $(-S_2)\fs{M}{m-\fs{\d^2}{2}}(1+\fs{S_2}{\beta})>1$,
\begin{align}\label{Fxs}
V(r,x)=\begin{cases}
-e^{-r}\fs{S_2}{\beta}\fs{M}{m-\fs{\d^2}{2}}\fs{(\beta+R_1)e^{R_1x}-(\beta+R_2)e^{R_2x}}{(R_1-S_2)e^{R_1b^*}-(R_2-S_2)e^{R_2b^*}},&x< b^*;\\
e^{-r}\left[\fs{M}{m-\fs{\d^2}{2}}+\fs{1}{S_2}e^{S_2(x-b^*)}\right],&x\ge b^*,
\end{cases}
\end{align}
where
$b^*=\fs{1}{R_1-R_2}\mbox{log}(\fs{R_2^2-S_2R_2}{R_1^2-S_2R_1})$.
And the optimal dividend strategy $L^*=\{l_s^*\}_{s\ge 0}$ is
\begin{align}\nonumber
l_s^*=M\mathbf{1}_{\{X_s^{L^*}\ge b^*\}},
\end{align}
Here $\mathbf{1}_{\{X_s^{L^*}>b^*\}}$ is the indicator function, which means that the optimal strategy is such that dividends are paid  at the maximum rate $M$ whenever $X_s^{L^*}\ge b^*$.
\end{thm}
$\mathbf{Proof}$ \;\;First, we show that $V(r,x)$ is a continuously  differentiable  solution of $(\ref{HJB})$.

 \noindent
If $(-S_2)\fs{M}{m-\fs{\d^2}{2}}(1+\fs{S_2}{\beta})\le 1$, denote $V(r,x)=e^{-r}F(x)$, where
\begin{align}\nonumber
F(x)=e^{-r}\fs{M}{m-\fs{\d^2}{2}}\left[1-e^{S_2x}(1+\fs{S_2}{\beta})\right].
\end{align}
Since $F(x)$ is a continuously  differentiable solution of equation (\ref{Fx}), it is easy to obtain that  $e^{-r}F(x)$ is a solution of (\ref{HJB}).
Similarly, if $(-S_2)\fs{M}{m-\fs{\d^2}{2}}(1+\fs{S_2}{\beta})>1$, denote $V(r,x)=e^{-r}F(x)$, where
\begin{align}
F(x)=\begin{cases}
-\fs{S_2}{\beta}\fs{M}{m-\fs{\d^2}{2}}\fs{(\beta+R_1)e^{R_1x}-(\beta+R_2)e^{R_2x}}{(R_1-S_2)e^{R_1b^*}-(R_2-S_2)e^{R_2b^*}},&x< b^*;\\
\fs{M}{m-\fs{\d^2}{2}}+\fs{1}{S_2}e^{S_2(x-b^*)},&x\ge b^*.
\end{cases}
\end{align}
From the fact that $F(x)$ is a continuously  differentiable solution of equation (\ref{Fx}), we obtain that $V(r,x)=e^{-r}F(x)$ is a solution of (\ref{HJB}).

From now on, we prove the optimality of strategy $L^*$.
Let $L$ be an admissible strategy with dividend rate $\{l_s\}_{s\ge 0}$. Let $\t^L$ denotes the ruin time of the surplus process. From the It$\hat{\mbox{o}}$ formula we obtain
\begin{align}\nonumber
E\left[V(r_{t\qx\t^L-}, X_{t\qx\t^L-})\right]=&\;V(r,x)+E\left[\jf_0^{t\qx\t^L}\left(-m V_r+\fs{\d^2}{2}V_{rr}+cV_x-l_sV_x\right)(r_{s-},X_{s-})ds\right]\\\nonumber
&\qquad+E\left[\sum_{0\le s<t\qx \t^L}(V(r_s,X_s)-V(r_s,X_{s-}))\right].
\end{align}
Thus, we obtain
\begin{align}\nonumber
V(r,x)=&-E\left[\jf_0^{t\qx\t^L}(-m V_r+\fs{\d^2}{2}V_{rr}+cV_x-l_sV_x)(r_{s-},X_{s-})ds\right]\\\nonumber
&\qquad-E\left[\sum_{0\le s\le t\qx \t^L}(V(r_s,X_s)-V(r_s,X_{s-}))\right]\\\nonumber=
&-E\bigg\{\jf_0^{t\qx\t^L}\bigg[(-m V_r+\fs{\d^2}{2}V_{rr}+cV_x-l_sV_x)(r_{s-},X_{s-})\\\nonumber
&\qquad-\l\jf_0^{X_{s-}}V(r_s,X_{s-}-y)\beta e^{-\beta y}dy+\l V(r_{s-},X_{s-})\bigg] ds\bigg\}\\\label{vef1}
\ge&\; E\left[\jf_0^{t\qx\t^L}e^{-r_s}l_sds\right].
\end{align}
We let $t\yjt\wq$ and use the dominated convergence theorem to get
\begin{align}
V(r,x)\ge E\left[\jf_0^{\t^L}e^{-r_s}l_sds\right].
\end{align}
If we use the  strategy $\{l_s^*\}_{s\ge 0}$, we get the equality in (\ref{vef1}) which leads to $V(r,x)=J^{L^*}(r,x)$. This completes the proof.\hfill$\Box$
\section{Ornstein-Uhlenbeck Process as a Interest  Rate}
In this section, we  consider  the Vasicek model as the  interest rate model. This model is based on the idea of mean-reversion, it tends to revert to a constant in the long run. This characteristic  can also be justified by economic arguments.  We refer the interested  readers to the  article of Vasicek\cite{vasicek1977equilibrium} for more details about the Vasicek model. The Vasicek model assumes the current short interest follows an Ornstein-Uhlenbeck process.
Denote  $\{r_s\}$ an Ornstein-Uhlenbeck process,  we can write it  as a stochastic differential equation of a standard Brownian motion
\begin{align}\label{30}
dr_s=a(\yj{b}-r_s)ds+\yj{\d} dB_s,
\end{align}
$a,\yj{\d},\yj{b}>0$ are constants. Here, $\yj{b}$ is the long-term mean of the process $\{r_s\}$, i.e. the interest rate process $\{r_s\}$ will evolve around $\yj{b}$ in the long run. The solution of the stochastic differential equation (\ref{30}) can be found by applying It$\hat{\bh{o}}$ lemma to $e^{at}r_t$, which leads to
\begin{align}\nonumber
r_s=re^{-as}+\yj{b}\left(1-e^{-as}\right)+\yj{\d} e^{-as}\jf_0^se^{au}dB_u,
\end{align}
with initial condition $r_0=r$.
Let $L=\{l_s\}_{s\ge 0}$ be an admissible strategy and $\t^L$ denotes the ruin time of surplus process $X_s^L$ with initial wealth $X_0=x$. The return function corresponding to $L$ is
\begin{align}
V^L(r,x)=E\left[\jf_0^{\t^L}e^{-\jf_0^sr_udu}l_sds\right],\;\;(r,x)\in \mathbb{R}\times \{\mathbb{R}^+\cup{0}\}.
\end{align}
It means the  dividend rate $l_s$ at time s  is discounted by the factor $e^{-\jf_0^sr_udu}$.
In the following, we write $U_s^y$ as $U_s=\jf_0^sr_udu$ with initial value $r_0=y$.
Our target is to maximize the expected discounted dividends given the preference rate $\{r_t\}$.
We define the value function as
\begin{align}\label{29}
V(r,x)=\sup_{L\in \mathcal{U}_{ad}}V^L(r,x),\;\;\;(r,x)\in \mathbb{R}\times \{\mathbb{R}^+\cup{0}\}.
\end{align}
The corresponding  Hamilton-Jacobi-Bellman equation is
\begin{align}\nonumber
\label{3}
&\left[-(r+\l)V+a(\yj{b}-r)V_r+\fs{\hat{\d}^2}{2}V_{rr}+cV_x\right](r,x)\\
&\quad+\l\jf_0^x V(r,x-y)dG(y)+\max_{0\le l\le M} l(1-V_x(r,x))=0.
\end{align}
Given a continuously differentiable function $\varphi(r,x):\mathbb{R}\times [0,+\wq)\yjt \mathbb{R}$, we define the operator
\begin{align}\nonumber
\mathcal{L}\left[\varphi\right]=&\left[-(r+\l)\varphi+a(\yj{b}-r)\varphi_r+\fs{\yj{\d}^2}{2}\varphi_{rr}+c\varphi_x\right](r,x)\\&+\l\jf_0^x \varphi(r,x-y)dG(y)+\max_{0\le l\le M} l\left(1-\varphi_x(r,x)\right).
\end{align}
This definition will make it easier for us to state the definition of viscosity solution.
\subsection{Properties of the Value Function}
In this subsection we prove the boundedness and   continuity of the value function $V$ which is defined in (\ref{29}).  The continuity makes it easier for us to define viscosity solution.
\begin{lem}
The value function $V$ is bounded.
\end{lem}
$\mathbf{Proof}$\;\;  Via Fubini's theorem, the value function satisfies
\begin{align}\label{Vdy}
V(r,x)&=\sup_{L\in \mathcal{U}_{ad}}V^L(r,x)
\le E\left[\jf_0^\wq e^{-U_s^r}Mds\right]
=E[\jf_0^\wq \mathbb{E}[e^{-U_s^r}]Mds].
\end{align}
Thanks to Borodin and Salminen (1998, p.525)\cite{borodin2012handbook}, we can use the fact that  $E[e^{-U_s^r}]=e^{f(r,s)}$, where
\begin{align}\label{frsqw}
f(r,s):=-\yj{b}s+\fs{\yj{\d}^2}{2\sigma^2}s-\fs{r-\yj{b}}{a}(1-e^{-as})+\fs{\yj{\d}^2}{4a^3}\big(1-(2-e^{-as})^2\big).
\end{align}
Let $b=\yj{b}-\fs{\yj{\d}^2}{2a^2}$  and $\tilde{\d}=\fs{\yj{\d}}{\sqrt{2a}}$. We can rewrite (\ref{frsqw}) as
\begin{align}\label{frs4}
f(r,s)=-bs-\fs{r-b}{a}(1-e^{-as})-\fs{\tilde{\d}^2}{2a^2}(1-e^{-as})^2.
\end{align}
Note that we can estimate the function $f$ as follows
\begin{align}\nonumber
f(r,s)&\ge -{b}s-\fs{\tilde{\d}^2}{2a^2}-\max\{\fs{r-b}{a},0\}.\\\label{frs2}
f(r,s)&\le -bs-\min\{\fs{r-b}{a},0\}.
\end{align}
From (\ref{Vdy}), (\ref{frs2}) and the assumption $b>0$, we obtain
\begin{align}\label{frs33}
V(r,x)\le ME\left[\jf_0^\wq e^{f(r,s)}ds\right]\le M\fs{e^{-\min\{\fs{r-b}{a},0\}}}{b}.
\end{align}
This shows that $V$ is bounded. $\hfill\Box$
\begin{rem}
We assume  $b>0$ because  it  helps  us to  obtain the boundedness of the value function.
\end{rem}
\begin{lem}
The value function is locally  Lipschitz continuous in $r$ and it is continuous  in $x$.
\end{lem}
$\mathbf{Proof}$ \quad The value function $V$ is obviously strictly increasing in $x$ and decreasing in $r$. Let $h\in \mathbb{R}^+$, $r\in \mathbb{R}$ and $L$ be an $\varepsilon-$optimal strategy for the initial point $(r,x)$. Then $L=\{l_t\}_{t\ge 0}$ is also an admissible strategy for $(r+h, x)$. In particular, $X^L$ denotes the wealth process with control strategy $L$, and $\t^L$ denotes the  time of ruin of the surplus process $X^L$. Therefore, one has
\begin{align}\nonumber
0&\ge V(r+h,x)-V(r,x)\\\nonumber
&\ge \mathbb{E}\left[\jf_0^{\t^L}e^{-U_s^{r+h}}l_sds-\jf_0^{\t^L}e^{-U_s^r}l_sds\right]-\varepsilon\\\nonumber
&=\mathbb{E}\left[\jf_0^{\t^L}e^{-U_s^r}l_s\left(e^{-\fs{h}{a}\left(1-e^{-as}\right)}-1\right)ds\right]-\e.
\end{align}
Using the fact that for all $s,$ $e^{-\fs{h}{a}\left(1-e^{-as}\right)}-1\ge \fs{h}{a}\left(e^{-as}-1\right)$ holds, we can see
\begin{align}\nonumber
0\ge V(r+h,x)-V(r,x)
\ge \fs{h}{a}\mathbb{E}\left[\jf_0^{\t^L} e^{-U_s^r}l_s\left(e^{-as}-1\right)ds\right]-\e.
\end{align}
From $e^{-as}-1\ge -1$, we can see
\begin{align}\nonumber
0&\ge V(r+h,x)-V(r,x)\\\nonumber
&\ge -\fs{h}{a}\mathbb{E}\left[\jf_0^{\t^L} e^{-U_s^r}l_sds\right]-\e\\\nonumber
&\ge-\fs{h}{a}V(r,x)-\e\\\nonumber
&\ge-hM\fs{e^{-\min\left\{\fs{r-b}{a},0\right\}}}{ab}-\e.
\end{align}
Here, in the last step, we used the fact that inequality (\ref{frs33}) holds.
This shows that $V$ is locally Lipschitz in $r$.

Now let $L$ be an $\e-$optimal strategy for the initial point $(r,x+h)$, with a slight abuse of notation $\t^L$, the ruin time of surplus process  $X^L$ with initial value $x+h$ is denoted by $\t^L$. $T_1$ denotes the first claim time of compound Poisson process. Define  $\t=\inf\{t\ge 0|\tilde{X_t}\bsy[0,x+h),\tilde{X}_0=x\}$, where $\tilde{X_t}$ denotes the surplus process driven by $\tilde{L}$ with initial value $x$.  Define $\tilde{L}=\{{\tilde{l}_t}\}_{t\ge 0}$ to be
\bea
\label{8}
\tilde{l}_t=
\ksfl
0, &t\le \t,\\
l_{t-\t}, &t>\t \;\bh{and} \;\tilde{X}_{\t}=x+h.
\jsfl
\eea
Strategy $\yw{L}$ means that $\yw{X}_t$ will not pay dividend until $\yw{X}_t$ attains $x+h$.
From now on, denote $\underline{h}=\fs{h}{c}$ for  simplicity. Then
\begin{align}
\nonumber
0&\;\le V(r,x+h)-V(r,x)\\\nonumber
&\;\le   V^L(r,x+h)+\e-\mathbb{E}\left[e^{-U_\t^r}\mathbf{1}_{\tilde{X}_\t=x+h}\jf_0^{\t^L}\exp\left\{{-U_s^{r_\t}}\right\}l_sds\right]\\\nonumber
&\;= V^L(r,x+h)+\e-\mathbb{E}\left[e^{-U_\t^r}\mathbf{1}_{\tilde{X}_\t=x+h}\jf_0^{\t^L}\exp\left\{{-\fs{1}{a}(r_\t-r)(1-e^{-as})}\right\}e^{-U_s^{r}}l_sds\right]
\end{align}
Since on $\{T_1\ge \fs{h}{c}\}$, there is no claims  between time $0$ and time $\underline{h}$, thus surplus process $\yw{X}_t$ attains $x+h$ at time $\underline{h}$, i.e. $\t=\underline{h}$ on $\{T_1\ge \fs{h}{c}\}$. We can obtain
\begin{align}
\nonumber
0&\;\le V(r,x+h)-V(r,x)\\\nonumber
&\;\le V^L(r,x+h)+\e-\mathbb{E}\left[e^{-U_{\underline{h}}^r}\mathbf{1}_{T_1\ge {\underline{h}}}\jf_0^{\t^L}\exp\left\{{-\fs{1}{a}(r_{\underline{h}}-r)(1-e^{-as})}\right\}e^{-{U}_s^r}l_sds\right]\\\nonumber
&\;=V^L(r,x+h)-\mathbb{E}\left[e^{-U_{\underline{h}}^r}
\mathbf{1}_{T_1\ge {\underline{h}}}\mathbf{1}_{\{r_{\underline{h}}\ge r\}}\jf_0^{\t^L}\exp\left[{-\fs{1}{a}(r_{\underline{h}}-r)(1-e^{-as})}\right]e^{-{U}_s^r}l_sds\right]\\\nonumber
&\quad\quad-\mathbb{E}\left[e^{-U_{\underline{h}}^r}\mathbf{1}_{T_1\ge \underline{h}}\mathbf{1}_{\{r_{\underline{h}}< r\}}\jf_0^{\t^L}\exp\left[{-\fs{1}{a}(r_{\underline{h}}-r)(1-e^{-as})}\right]e^{-{U}_s^r}l_sds\right]+\e.
\end{align}
From the fact that $T_1$ is independent of $\{r_t\}$, we can deduce that
\begin{align}
\nonumber
0&\;\le V(r,x+h)-V(r,x)\\\nonumber
&\;\le V^L(r,x+h) -E\left[e^{-U_{\underline{h}}^r}\mathbf{1}_{\{r_{\underline{h}}\ge r\}}\jf_0^{\t^L}(1+\fs{1}{a}(r-r_{\underline{h}})(1-e^{-as}))e^{-{U}_s^r}l_sds\right]e^{-\l\underline{h}} \\\nonumber
&\;\quad\quad-E\left[e^{-U_{\underline{h}}^r}\mathbf{1}_{\{r_{\underline{h}}<r\}}\jf_0^{\t^L}e^{-U_s^r}l_sds\right]e^{-\l\underline{h}}+\e\\\label{conx}
&\;\le V^L(r,x+h)\left[1-E[e^{-U_{\underline{h}}^r}
]e^{-\l\underline{h}}\right]+V^L(r,x+h)E\left[e^{-U_{{\underline{h}}}^r}\mathbf{1}_{{\{r_{\underline{h}}>r\}}}
(r_{\underline{h}}-r)\right]\fs{1}{a}e^{-\l\underline{h}}+\e.
\end{align}
In Borodin and Salminen(1998, p525)\cite{borodin2012handbook}, we can find the distribution of $\exp\{{-U_{\underline{h}}^r}\}$ and $r_{\underline{h}} \exp\{{-U_{\underline{h}}^r}\}$. Calculating  the expectation in the square brackets directly, we find that there exists a constant $Q_1$ such that,  for $h$ small enough, we have
\bea\label{q1}
E\left[e^{-U_{\underline{h}}^r}\mathbf{1}_{{\{r_{\underline{h}}>r\}}}(r_{\underline{h}}-r)\right]\fs{1}{a}e^{-\l\underline{h}}\le Q_1\sqrt{h}.
\eea
And there also  exists a constants $Q_2$ such that
\bea\label{q2}
1-E\left[e^{-U_{\underline{h}}^r}\right]e^{-\l\underline{h}}\le Q_2 h.
\eea
Substituting (\ref{q1}) and (\ref{q2}) into (\ref{conx}), we obtain that  there exists  a constant $Q$ such that
\bea\nonumber
0\le V(r,x+h)-V(r,x)\le V(r,x+h)Q\sqrt{h}+\e\le \fs{Me^{-\min\left\{\fs{r-b}{a},0\right\}}}{b}Q\sqrt{h}+\varepsilon.
\eea
This proves the continuity of the value function.  $\hfill\Box$

We do want to explore more regularity properties about the value function, but unfortunately, in many applications the value function $V(r,x)$ is not necessarily smooth, or it can be very difficult to prove its differentiability. Therefore we need to introduce the notation of weak solutions, namely viscosity solutions.

We recall that the notion of viscosity solutions was introduced by Crandall and Lions \cite{crandall1983viscosity} for the first order equations and Lions \cite{lions1983optimal,lions1983optimal2} for the second order equations. The notion of viscosity solution of integro-differential equations was pursued by Soner\cite{soner1986optimal}. The viscosity solution concept  of  fully nonlinear partial differential equations has been proving to be extremely useful for   control theory  due to the fact that it does not need the differentiability of the value function.
 It  merely  requires  continuity of the value function to  define the viscosity  solution. We refer to the user's guide of Crandall, Ishii and Lions\cite{crandall1992user} for an overview of the theory of viscosity solutions and their applications. Using the notion of viscosity solution we   prove that the value function is the (viscosity) solution of the corresponding equation (\ref{3}).  The viscosity solution approach is becoming a well established approach to study stochastic control  problem, see, e.g. the books \cite{fleming2006controlled,yong1999stochastic}.
\begin{defn}\label{defsub}
We say that a continuous function $\underline{u}: \mathbb{R}\times [0,\wq)\yjt \mathbb{R}$ is a viscosity subsolution of (\ref{3}) at $(r,x)\in \mathbb{R}\times \mathbb{R}^+$ if any  continuously differentiable function $\vf:\mathbb{R}\times (0,\wq)\yjt \mathbb{R}$
with $\vf(r,x)=\underline{u}(r,x)$ such that  $\underline{u}-\vf$ reaches the maximum at $(r,x)$ satisfies
 \begin{align}\nonumber
 \mathcal{L}[\vf](r,x)\ge 0.
 \end{align}
\label{defsup}
We say that a continuous function $\bar{u}: \mathbb{R}\times [0,\wq)\yjt \mathbb{R}$ is a viscosity supersolution of (\ref{3}) at $(r,x)\in \mathbb{R}\times \mathbb{R}^+$ if any continuously differentiable function $\vf:\mathbb{R}\times (0,\wq)\yjt \mathbb{R}$ with $\vf(r,x)=\bar{u}(r,x)$ such that $\bar{u}-\vf$ reaches the minimum  at $(r,x)$ satisfies
\begin{align}\nonumber
 \mathcal{L}[\vf](r,x)\le 0.
\end{align}
Finally, we call a continuous function $u:\mathbb{R}\times[0,\wq)\yjt \mathbb{R}$ is a viscosity solution of (\ref{3}) if it is both a viscosity subsolution and a viscosity supersolution at any $(r,x)\in\mathbb{R}\times \mathbb{R}^+$.
\end{defn}
\begin{thm}
The value function $V$ defined in (\ref{29}) is a viscosity solution of (\ref{3}) on $(0,+\wq)$.
\end{thm}
$\mathbf{Proof}$\;\;
First, we  show that the value function is a viscosity supersolution of (\ref{3}).  Here we  claim that the dynamic programming principle holds: i.e., for any  $(r,x)\in \mathbb{R}\times [0,+\wq)$  and any stopping time $\t$, we have
\begin{align}\label{DPP}
V(r,x)=\sup_{L\in \hx{U}_{ad}}E\left[\jf_0^{\t\qx\t^L}e^{-\jf_0^s r_udu}l_sds+e^{-\jf_0^{\t\qx\t^L}r_udu}V(r_{\t\qx\t^L},X_{\t\qx\t^L})\right].
\end{align}
This principle can be proving  by similar methods from Azcue and Muler \cite{azcue2005optimal}. We consider the following strategy: The company always pays dividends at rate $l_0$ until time of ruin, where  $l_0\in [0,M]$ is a positive constant. Let $X_t^0$ denotes the surplus  process controlled by strategy $l_0$.
 Denote $\t_1$ the first claim time of the surplus process. Let $\f$ be a  continuously differentiable function on $\mathbb{R}\times [0,+\wq)$ such that  $V-\f$ attains its minimum 0 at $(r,x)$. By the dynamic programming principle,
 we get
\begin{align}
\nonumber
0&\;\ge l_0 E\left[\jf_0^{\t_1\qx h}e^{-U_s^r}ds\right]+E\left[e^{-U_{\t_1\qx h}^r}V(r_{\t_1\qx h}, X_{\t_1\qx h}^0)\right]-V(r,x)\\
\nonumber
&\;\ge l_0 E\left[\jf_0^{\t_1\qx h}e^{f(r,s)}ds\right]+E\left[e^{-U_{\t_1\qx h}^r}\f(r_{\t_1\qx h}, X_{\t_1\qx h}^0)\right]-\f(r,x)\\
\nonumber
&\;= l_0 E\left[\jf_0^{\t_1\qx h}e^{f(r,s)}ds\right]+E\left[e^{-U_{\t_1\qx h}^r}\left[\f(r_{\t_1\qx h}, X_{\t_1\qx h}^0)-\f(r_{\t_1\qx h-}, X_{\t_1\qx h-}^0)\right]\mathbf{1}_{\{\t_1<h\}}\right]\\\nonumber
&\;\quad\quad+ E\left[e^{-U_{\t_1\qx h}^r}\f(r_{\t_1\qx h-}, X_{\t_1\qx h-}^0)-\f(r,x) \right]:=I_1+I_2+I_3.
\end{align}
where $I_i, i = 1, 2, 3$ are the three terms on the right hand side above. Clearly, we have
\begin{align}
\nonumber
I_1&\;=l_0E\left[\jf_0^{\t_1\qx h}e^{f(r,t)}dt\right]=l_0E\left[\jf_0^h \mathbf{1}_{\{\t_1\ge t\}} e^{f(r,t)}dt \right]=l_0 \jf_0^h e^{-\l t}e^{f(r,t)}dt,\\
\nonumber
I_2&\;= E\left[\jf_0^h\l e^{-\l t}\jf_0^{X_{t-}^0} e^{-\jf_0^tr_sds}\big\{\f(r_t,X_{t-}^0-y)-\f(r_t,X_{t-}^0)\big\}dG(y)dt \right],\\
\nonumber
I_3&\;= E\bigg[\jf_0^{h}\mathbf{1}_{\{\t_1\ge t\}} e^{-\jf_0^tr_sds}\big[-r_t\f(r_t,X_{t-}^0)+a(\yj{b}-r_t)\f_r(r_t,X_{t-}^0)+\fs{{\yj{\d}^2}}{2}\f_{rr}(r_t,X_{t-}^0)\\\nonumber
&\quad+c\f_x(r_t,X_{t-}^0)-l_0\f_x(r_t,X_{t-}^0)\big]dt  \bigg].
\end{align}
Let us  sum those three   together and divide  by $h$. Letting $h\yjt 0$ and using the fact that $l_0$ is arbitrary, we obtain
\bea \nonumber
\mathcal{L}\f(r,x)\le 0.
\eea
This proves that  the value function is  a viscosity supersolution of equation (\ref{3}).

Now we prove that the value function is a viscosity subsolution of the corresponding HJB equation. Assume the contrary, i.e.  there exists a point $(r_0,x_0)\in \mathbb{R}\times \mathbb{R}^+$ such that $V$ is not a viscosity subsolution. By the definition of viscosity solution, there exists $\eta>0$ and a continuously differentiable function $\vf^0$ such that $V(r_0,x_0)=\vf^0(r_0,x_0)$, $\vf^0(r,x)\ge V(r,x)$ on  $\mathbb{R}\times \mathbb{R}^+$ and
\begin{align}\nonumber
\mathcal{L}[\vf^0](r_0,x_0)=-2\eta<0.
 \end{align}
First, we assume that $r_0\ge 0$ ($r_0<0$ can be proved similarly).  Consider  the  function
\begin{align}\label{ou1}
\yj{\vf}(r,x)=\vf^0(r,x)+\fs{\eta}{x_0^2\l}(x-x_0)^2+\fs{\eta}{\l}(r-r_0)^4,
\end{align}
then we can notice  that $\yj{\vf}(r_0,x_0)=\vf^0(r_0,x_0),\yj{\vf}_x(r_0,x_0)=\vf_x^0(r_0,x_0), \yj{\vf}_{r}(r_0,x_0)=\vf_r^0(r_0,x_0)$,
$\yj{\vf}_{rr}(r_0,x_0)=\vf_{rr}^0(r_0,x_0)$, and
\begin{align}\nonumber
\l\jf_0^{x_0}\yj{\vf}(r_0,x_0-y)dG(y)=\;&\l\jf_0^{x_0}\left[\vf^0(r_0,x_0-y)+\fs{\eta}{x_0^2\l}y^2\right]dG(y)\\\nonumber
\le\;& \l\jf_0^{x_0}\vf^0(r_0,x_0-y)dG(y)+\eta.
\end{align}
We can get
\begin{align}\nonumber
\mathcal{L}[\yj{\vf}](r_0,x_0)\le-\eta<0.
\end{align}
Since $\yj{\vf}$ is nonnegative and continuously differentiable, we can find  $h\in(0,\fs{x}{2})$ such that
\begin{align}\label{ou4}
\mathcal{L}[\yj{\vf}](r,x)\le-\fs{\eta}{2}<0
\end{align}
on $(r,x)\in [r_0-2h,r_0+2h]\times[x_0-2h,x_0+2h]$.
Let $\psi$ be an even and  nonnegative continuously differentiable function with support included in $(-1,1)\times(-1,1)$ such that
$\jf_{-1}^1\jf_{-1}^1\p(r,y)drdy=1$. We define $\nu_n:(-\wq,\wq)\times[0,\wq)\yjt\mathbb{ R}$ as the convolution
\begin{align}
\nu_n(r,y)=\fs{1}{n^2}\jf\jf_{\sqrt{|y-x|^2+|r-s|^2}<\fs{1}{n}}\p(n(r-s),n(y-x))\left(V(s,x)+\fs{\eta h^2}{2\l x_0^2}+\fs{\eta h^4}{2\l}\right)dsdx.
\end{align}
Since $V$ is not defined on $ \mathbb{R}\times\mathbb{R}^-$ in this integral, we can extend $V$ as $V(r,y)=V(r,0)+y$ for $(r,y)\in\mathbb{R}\times \mathbb{R}^-$. By standard techniques (e.g., see Wheeden and Zygmund \cite{zhyg}), we have that $\nu_n$ is a smooth function and $\nu_n$ converges to $V+\fs{\eta h^2}{2\l x_0^2}+\fs{\eta h^4}{2\l}$ uniformly on $[r_0-2h,r_0+2h]\times[0,x+h]$. Then,
  we can find  $n_0$ large enough such that
\begin{align}\label{ou2}
V(r,y)+\fs{\eta h^2}{\l x_0^2}+\fs{\eta h^4}{\l}\;\ge \nu_{n_0}(r,y)\;\ge V(r,y)+\fs{\eta h^2}{4\l x_0^2}+\fs{\eta h^4}{4\l}.
\end{align}
Let $\chi$ be a continuously differentiable function satisfying the following conditions
\begin{enumerate}[(1)]
\item $0\le \chi\le 1 $,
\item $\chi(r,y)=1$ for $(r,y)\in[r_0-h,r_0+h]\times[x_0-h,x_0+h]$,
\item $\chi(r,y)=0$ for $(r,y)\bsy [r_0-2h,r_0+2h]\times[x_0-2h,x_0+2h]$.
\end{enumerate}
Define the  function
\begin{align}\label{ou3}
\vf(r,y)=\;\chi(r,y)\yj{\vf}(r,y)+(1-\chi(r,y))\nu_{n_0}(r,y).
\end{align}
Take $\varepsilon=\min\left\{\fs{\eta}{2(r_0+h)},\fs{\eta h^4}{\l},\fs{\eta h^2}{4\l x_0^2}\right\}$, from (\ref{ou1}), (\ref{ou2}), (\ref{ou3}) we can see that function $\vf(r,y)$ satisfies
\begin{align}\label{ou6}
[V-\vf](r,y)\le -\varepsilon
\end{align}
on $\{r_0-h\}\times[x_0-h,x_0+h]\cup\{r_0+h\}\times[x_0-h,x_0+h]\cup[r_0-h,r_0+h]\times[0,x_0-h]\cup
[r_0-h,r_0+h]\times\{x_0+h\}$.
From (\ref{ou4}), we obtain
\begin{align}\label{ou5}
\mathcal{L}[\vf](r,y)\le -r\varepsilon
\end{align}
on $[r_0-h,r_0+h]\times[x_0-h,x_0+h]$.
For any  strategy  $L=\{l_t\}_{t\ge 0}$, denote
\begin{align}\nonumber
&\bar{\t}=\inf\left\{t>0:X_t^L\ge x_0+h \; \mbox{or}\; r_t\notin[r_0-h,r_0+h]\right\},\\\nonumber
&\underline{\t}=\inf\left\{t>0:X_t^L\le x_0-h \;\mbox{or}\; r_t\notin[r_0-h,r_0+h]\right \}.
\end{align}
Take $\t=\bar{\t}\qx\underline{\t} $.
Since $\vf$ is continuously  differentiable, we can see that
\begin{align}\nonumber
&\;E\left[\vf(X_\t,r_\t)e^{-\jf_0^\t r_sds}\right]-\vf(r_0,x_0)\\\nonumber
=&\; E\bigg\{\jf_0^\t e^{-\jf_0^ur_sds}\left[a(\yj{b}-r_u)\vf_r-r_u\vf-l_u\vf_x+c\vf_x+\fs{1}{2}\yj{\d}^2\vf_{rr}\right](r_u,X_{u-})du\\\nonumber
&\;\quad+\jf_0^\t e^{-\jf_0^ur_sds}\left[\l\jf_0^{X_{u-}}\vf(r_u,X_{u-}-y)dG(y)-\l\vf(r_u,X_{u-})
\right]du\bigg\}\\\nonumber
\le&\; E\left[ \jf_0^\t e^{-\jf_0^ur_sds}\mathcal{L}[\vf](r_u,X_{u-})du-\jf_0^\t e^{-\jf_0^ur_sds}l_udu\right]\\\nonumber
\le &\; -\varepsilon E\left[ \jf_0^\t e^{-\jf_0^ur_sds}r_udu\right]-E\left[\jf_0^\t e^{-\jf_0^ur_sds}l_udu\right].
\end{align}
The last inequality holds because of (\ref{ou5}).
Combining with (\ref{ou6}), we can see
\begin{align}\nonumber
&\;E\left[e^{-\jf_0^\t r_sds} V(r_\t,X_\t)\right]\\\nonumber
\le &\; E\left[e^{-\jf_0^\t r_sds}(\vf(r_\t,x_\t)-\varepsilon) \right]\\\nonumber
=&\;E\left[e^{-\jf_0^\t r_sds}\vf(r_\t,x_\t)-\vf(r_0,x_0)\right]+E\left[\vf(r_0,x_0)-e^{-\jf_0^\t r_sds}\varepsilon\right]\\\nonumber
\le& \;-\varepsilon E\left[ \jf_0^\t e^{-\jf_0^ur_sds}r_udu\right]-E\left[\jf_0^\t e^{-\jf_0^ur_sds}l_udu\right]+E\left[\vf(r_0,x_0)-e^{-\jf_0^\t r_sds}\varepsilon\right].
\end{align}
Since
\begin{align}\nonumber
E\left[ \jf_0^\t e^{-\jf_0^ur_sds}r_udu\right]=1-E\left[e^{-\jf_0^\t r_sds}\right],
\end{align}
we obtain
\begin{align}\nonumber
&\;E\left[e^{-\jf_0^\t r_sds} V(r_\t,X_\t)\right]\\\nonumber
\le\; &\vf(r_0,x_0)-\varepsilon-E\left[\jf_0^\t e^{-\jf_0^ur_sds}l_udu\right]\\\nonumber
=&\;V(r_0,x_0)-\varepsilon-E\left[\jf_0^\t e^{-\jf_0^ur_sds}l_udu\right].
\end{align}
Since strategy $L$ is arbitrary, using the Dynamic Programming Principle (\ref{DPP}), we can see that
\begin{align}\nonumber
V(r_0,x_0)=\sup_{L\in \mathcal{U}_{ad}}E\left[\jf_0^\t e^{-\jf_0^ur_sds}l_udu +e^{-\jf_0^\t r_sds} V(r_\t,X_\t)\right]\le V(r_0,x_0)-\varepsilon.
\end{align}
This is a contradiction. This shows that the value function is also a viscosity subsolution of (\ref{3}). $\hfill\Box$
\section{Concluding Remarks}
In this paper we investigate the optimal dividend of  insurance company under the assumption of stochastic interest rate   and give out the explicit expression of the optimal strategy when the interest rate follows a geometric Brownian motion and the claim sizes follow the exponential distribution. For the case of the Vasicek model, we did not give out the solution of the  value function but we explored its properties  and we used the notion of viscosity solution to create the connection between the value function and the HJB equation, which is important for the future study about the optimal strategy.

When the discounting factor is given by a geometric Brownian motion, we can see that the optimal strategy is still a threshold strategy, except some changes in the parameters compared with the case of deterministic interest rate. This partly used the fact that the surplus process is independent of the discounting factor, which provides a convenient condition for us  to prove the optimality. Only  exponential claims are considered in section 3, but we already started to explore more general cases of claim distributions. We  conjecture that in the setting of geometric Brownian motion, the optimal dividend is a band strategy if the claim follows a more general continuous distribution function $G(y)$.

In section 4, we consider the dividend maximization problem when stochastic interest rate follows an Ornstein-Uhlenbeck Process. But we do not give out more regularity properties of the  value function. It is quite hard to find an  explicit expression of the dividend strategy. We will focus on comparison principle and  optimal strategy in future research.

\medskip
\noindent$\mathbf{Acknowledgements}$\;\;
This work is supported by the NSF of China (No. 11471171 and No. 11571189).  Here we want to express our thanks to  Lihua Bai and Junyi Guo for their valuable insights and suggestions. Thanks to Jacques Rioux for his  dedication to the improvement of this paper.

\end{document}